\documentclass {elsart}
\usepackage{graphicx}

\usepackage{graphics}
\usepackage{graphicx}
\usepackage{epsfig}

\usepackage{amssymb}

\begin{document}

\begin{frontmatter}



\title{Hellmann-Feynman connection for the relative Fisher information}


\author[SRC]{R. C. Venkatesan\corauthref{cor}}
\corauth[cor]{Corresponding author.}
\ead{ravi@systemsresearchcorp.com;ravicv@eth.net}
\author[UNLP]{A. Plastino}
\ead{plastino@fisica.unlp.edu.ar}

\address[SRC]{Systems Research Corporation,
Aundh, Pune 411007, India}
\address[UNLP]{IFLP, National University La Plata \&
National Research   (CONICET)\\ C. C., 727 1900, La Plata,
Argentina}

\begin{abstract}
The $(i)$ reciprocity relations for the relative  Fisher information (RFI,
hereafter) and $(ii)$ a generalized RFI-Euler theorem, are self-consistently derived from the Hellmann-Feynman theorem.
  These new
reciprocity relations  generalize the RFI-Euler theorem and constitute
the basis for building up  a mathematical Legendre transform structure
(LTS, hereafter), akin to that of thermodynamics, that
underlies the RFI scenario. This demonstrates the possibility of translating
the entire mathematical structure of thermodynamics into a  RFI-based
theoretical framework. Virial theorems play a prominent role in this endeavor,
as a Schr\"{o}dinger-like equation  can be associated to the RFI.   Lagrange
multipliers are determined invoking the RFI-LTS link and the quantum
mechanical virial theorem. An appropriate ansatz allows for the
inference of probability density functions (pdf's, hereafter) and
energy-eigenvalues of the above mentioned Schr\"{o}dinger-like equation.
The energy-eigenvalues obtained here via inference are
benchmarked against established theoretical and numerical results.  A principled theoretical basis to reconstruct the RFI-framework from the FIM framework is established.
Numerical examples for exemplary cases are provided.
\end{abstract}

\begin{keyword}
Relative Fisher information \sep Hellmann-Feynman theorem\sep reciprocity relations \sep generalized RFI-Euler theorem
\sep
Legendre transform structure \sep energy-eigenvalue inference.

PACS: 05.20.-y; \ 2.50.Tt; \ 03.65.-w
\end{keyword}
\end{frontmatter}
\section{Introduction}
The relative Fisher information (RFI, hereafter)  is a measure of
uncertainty that is the focus of much attention in statistical
physics, estimation theory, and allied disciplines (see [1]).  The RFI is defined by [2, 3]
\begin{equation}
\Im \left[f|g \right] =
\int_{\Re ^n } {f\left( \textbf{x} \right)
\left| {\nabla \ln \frac{{f\left( \textbf{x} \right)}}{{g\left( \textbf{x} \right)}}} \right|} ^2 d\textbf{x},
\end{equation}
where $\left|  \bullet  \right|^2$ is the square norm.  Here,
$g(\textbf{x})$ is the \textit{reference probability} which may be
construed as encapsulating \textit{prior knowledge}.  Some of the
various forms in which the RFI has been expressed have been cited
in Refs. [4-11].  It is noteworthy to mention that a form of the RFI
known as the 'weighted Fisher information' has been
self-consistently derived on the basis of estimation theory in
[12]. This also allowed the class of efficient pdf's $f(\textbf{x})$ to be computed. These achieve minimizing the \textit{weighted} mean-squared error of estimation.  

The RFI may be legitimately construed as relating to the Fisher information measure (FIM, hereafter) akin to the manner in which the Kullback-Leibler divergence (K-Ld, hereafter) relates to the Shannon entropy [13].  Specifically, both the RFI and the K-Ld are not only measures of uncertainty, but also measures of discrepancy between two probability distributions.
However, while the derivative term in the RFI tacitly implies its "localization" or "fine-grained" attributes,
 the K-Ld is, comparatively, "coarse-grained".  Ref. [1] demonstrated that the mathematical structure of
thermodynamics seamlessly translates into an RFI context,
within the framework of a time independent model.  This was
accomplished via the derivation of a generalized RFI-Euler theorem
(of the form specified in [14]) and of the concomitant Legendre
transform structure (LTS, hereafter) for the RFI.  The basis for
this LTS-derivation  is a one dimensional time independent
Schr\"{o}dinger-like equation. One speaks of  the
Schr\"{o}dinger-like link for the RFI (S-RFI link, hereafter).   This S-RFI link
 is obtained by specifying the \textit{reference
probability} $g(x)$ in (1) for the one dimensional case (eg. [4])
\begin{equation}
 g\left( x \right) = e^{ - V\left( x \right)}; \int {e^{ - V\left( x \right)} dx}  = 1, \\
\end{equation}
known as the \textit{Gibbs form}, where $V(x)$  is an a-priori
specified \textit{convex} physical potential, such that $V_{xx}(x)>0$.
By Eqs. (1) and (2)
\begin{equation}
\Im \left[ {f\left| {e^{ - V\left( x \right)} } \right.} \right] =
\int {f\left( x \right)\left| {\nabla \left( {\ln f\left( x
\right) + V\left( x \right)} \right)} \right|^2 dx}.
\end{equation}
\textit{Note that the so-called Gibbs form in (2) is perfectly acceptable within
the context of this study, since equilibrium probabilities in quantum mechanics (QM, hereafter) are described
by an exponential form}.  Setting $f(x)=\psi^2(x)$ and performing
a variational extremization of (3) with respect to $\psi(x)$, subject to
constraints, yields [1]
\begin{equation}
\begin{array}{l}
 - \frac{1}{2}\frac{{d^2 \psi \left( x \right)}}{{dx^2 }}
  - U_{RFI}\left( x \right)\psi \left( x \right) =   \frac{{\lambda _0 }}{8}\psi \left( x \right),\\
 \\
where\\
\\
 U_{RFI}\left( x \right) = \frac{1}{8}\left[ \sum\limits_{i = 1}^M {\lambda _i A_i \left( x \right)}
  -  {V^2_x \left( x \right)}   + 2V_{xx} \left( x \right) \right],
 \end{array}
\end{equation}
 is a pseudo-potential.  Here, $\psi(x)$ is a real quantity (as always in a one dimensional setting  [15]),
 $V_x \left( x \right) = \frac{{dV\left( x \right)}}{{dx}}$,
  and $V_{xx} \left( x \right) = \frac{{d^2 V\left( x \right)}}{{dx^2 }}$.  The $\lambda_i$
  are  $M$ Lagrange multipliers associated to
   the expectation of the empirical observables $
\left\langle {A_i \left( x \right)} \right\rangle  = \int {A_i \left( x \right)\psi ^2 \left( x \right)dx}
$, and $\lambda_0$ is the Lagrange multiplier
for the normalization condition $\int {\psi ^2 \left( x \right)dx}  = 1$.
Ref. [1] also derived a procedure to infer the energy-eigenvalues of the S-RFI link,
 via solution of a linear partial differential equation (PDE)
\begin{equation}
\lambda _0  = \sum\limits_{i = 1}^M {\left( {1 + k} \right)\lambda _i \frac{{\partial \lambda _0 }}{{\partial \lambda _i }}},
\end{equation}
derived solely with recourse to the LTS of the RFI and the QM virial theorem.

The Hellmann-Feynman theorem (HFT, hereafter) [16, 17] plays  a
central role in the application of QM.  It demonstrates the relationship between
perturbations in an operator in inner product spaces and the
concomitant perturbations in the operator's eigenvalue. The HFT
states that a nondegenerate eigenvalue $E(\lambda)$ of a
parameter-dependent Hermitian operator $H(\lambda)$, with
associated (normalized) eigenvector $\psi(\lambda)$, relate as
\begin{equation}
\frac{{\partial E_i }}{{\partial \lambda }} = \left\langle {\psi _i \left( \lambda   \right)} \right|\frac{{\partial H}}{{\partial \lambda  }}\left|
{\psi _i \left( \lambda   \right)} \right\rangle.
\end{equation}

Earlier works
established a relation between the HFT and the LTS of FIM  [18, 19]. Refs. [18,19] make
substantial use of the LTS of the FIM derived in [20]. In the
case of FIM-based treatments, the empirical observables $A_i(x)$
are the only source of prior knowledge.

In RFI variational
extremizations (4), both the empirical observables $A_i(x)$ and
the convex physical potential $V(x)$ constitute prior knowledge.
 Thus, studies within the RFI framework will
 substantially differ, in a qualitative sense, from those obtained for the FIM scenario.
 Specifically, as it is evidenced by (4), values of $\psi(x)$ are determined by both
 the derivative terms of the physical potential $V(x)$ and the empirical contribution in (4), i.e. the $A_i(x)$.

 Before proceeding further, it is of paramount importance to clarify certain
 terminologies and definitions.  In this paper, the RFI is the primary measure of uncertainty.
  Likewise, the term FIM framework is used to described treatments for which the FIM is the primary measure of uncertainty
  (see Refs. [18, 19]).  The expression for the RFI can be expressed in terms of a FIM
  and expectations of derivative terms of the physical potential $V(x)$,
   which comprise the reference probability $g(x)$ in (1)-(3).
   \textit{However, the pdf in (3) and (11) is the one which extremizes the RFI and not the FIM}.

The goals of this paper are \begin{itemize}

\item $(i)$ to establish that the
reciprocity relations and the generalized RFI-Euler theorem which
are central to the LTS of the RFI [1] may be derived from
first-principles from  the HFT (Section 3), by specializing
\textit{virial theorems} [21] to the information-theoretic domain,

\item $(ii)$ to establish with the aid of the HFT, a principled basis to
describe the interplay between the probability amplitude
$\psi(x)$, the derivatives of the \textit{a-priori} specified
convex physical potential $V(x)$, and the empirical observables
$A_i(x)$.  This is accomplished in Sections 4 by inference of an
ansatz for the pdf,

\item $(iii)$ to demonstrate  the inference of the energy-eigenvalues of the
S-RFI link, by recourse only to the QM virial
 theorem and the reciprocity relations.   This is achieved by subjecting (5)
  to the reciprocity relation (8) in Section 2, thereby demonstrating
  the practical utility of (5) in actual inference settings.
  The said task is accomplished in Section 5 by inference of pdf's whose profile
  is dictated by the ansatz introduced in Section 4,

\item $(iv)$  to exemplify the distinct nature of the pdf profiles inferred from empirical
observables within the RFI framework vis-\'{a}-vis equivalent Fisher-based ones.
This is accomplished in Section 5.
\textit{Notably, Section 5 establishes the fact that solutions
 of the variational extremization of the FIM constitute a subset of solutions within the RFI framework}.
  Specifically, it is established that
   the Lagrange multipliers $\lambda_k$ associated with the constraints in variational extremizations performed in both
   the RFI and the FIM frameworks are identical.  This is attributed to the \textit{Gibbs form} employed
   in Eqs. (2) and (3).  Section 5 also demonstrates
   that the energy-eigenvalues for the cases of the harmonic oscillator
   potential and the quartic anharmonic oscillator potential, inferred from empirical observations,
    show remarkable consistency with established studies [22,23],
\item $(v)$ a theoretical basis to reconstruct the pdf of the RFI framework from that of the FIM-case is established. \textit{This reconstruction demonstrates the efficacy of the relation (13) in inference scenarios}.  The fact that the Lagrange multipliers associated with the constraints in the RFI-framework and the FIM-framework are equal - a feature of the \textit{Gibbs form} employed in (2) and (3), is shown to ameliorate the process of reconstruction.  \end{itemize}

Numerical examples for exemplary
cases are provided.  To the best of the authors' knowledge, these objectives have never hitherto been accomplished.

\section{Theoretical preliminaries}

The RFI-Legendre transform structure
 links the RFI-measure, the normalization Lagrange multiplier $\lambda_0$,
  the prior information contained in a set of expectation values, and the Lagrange multipliers related to them.
 Throughout this paper, unless specifically specified otherwise,
    expectations are denoted by $<\bullet>_{\psi^2_{RFI}(x)}$ and are evaluated with respect to the pdf
    $f_{RFI}(x)=\psi^2_{RFI}(x)$ which extremizes the RFI subject to constraints. (Cf. Eq. (4)).  It is therefore desirable to define the RFI as $
\Im \left[ {\psi ^2_{RFI} \left( x \right)\left| {e^{ - V\left( x \right)} } \right.} \right]$, instead of $
\Im \left[ {\psi \left( x \right)\left| {e^{{{ - V\left( x \right)} \mathord{\left/
 {\vphantom {{ - V\left( x \right)} 2}} \right.
 \kern-\nulldelimiterspace} 2}} } \right.} \right]$ (as is the case in [1]). The pertinent relationships from  [1] remain unaffected, and are re-stated as
\begin{equation}
\lambda _0 \left( {\lambda _1 ,.,\lambda _M } \right) = \Im \left( {\left\langle {A_1(x) } \right\rangle_{\psi ^2_{RFI}} ,.,\left\langle {A_M(x) } \right\rangle_{\psi ^2_{RFI}}
}\right) - \sum\limits_{i = 1}^M {\lambda _i \left\langle {A_i(x) } \right\rangle_{\psi ^2_{RFI}} },
\end{equation}
plus the reciprocity relations and the generalized RFI-Euler theorem, i.e.,
\begin{equation}
\frac{{\partial \lambda _0 }}{{\partial \lambda _i }} =  - \left\langle {A_i(x) } \right\rangle_{\psi ^2_{RFI}},
\end{equation}

\begin{equation}
\frac{{\partial \Im \left[ {\psi^2_{RFI}(x) \left| {e^{ - V\left( x \right)} } \right.} \right]  }}{{\partial \left\langle {A_j(x) } \right\rangle_{\psi ^2_{RFI}} }} = \lambda
_j,
\end{equation}
and
\begin{equation}
\frac{{\partial \Im \left[ {\psi^2_{RFI}(x) \left| {e^{ - V\left( x \right)} } \right.} \right]}}{{\partial \lambda _i }} = \sum\limits_{j = 1}^M {\lambda _j
\frac{{\partial \left\langle {A_j(x) } \right\rangle_{\psi ^2_{RFI}} }}{{\partial \lambda _i }}},
\end{equation}
respectively.  The RFI obeys the following relations [1]
\begin{equation}
\Im \left[ {\psi^2_{RFI}(x)\left| {e^{ - V\left( x \right)} } \right.} \right] = I\left[ \psi^2_{RFI} \right] - 2\left\langle {V_{xx} \left( x \right)} \right\rangle_{\psi ^2_{RFI}}
+ \left\langle {V_x^2 \left( x \right)} \right\rangle_{\psi ^2_{RFI}}\\
\end{equation}
and
\begin{equation}
\Im \left[ {\psi^2_{RFI}(x)\left| {e^{ - V\left( x \right)} } \right.} \right]=\lambda _0  + \sum\limits_{i = 1}^M {\lambda _i \left\langle {A_i \left( x
\right)} \right\rangle_{\psi ^2_{RFI}} },
\end{equation}
where $I[\psi^2_{RFI}]$ is the FIM defined in terms of the amplitude  $\psi_{RFI}$.
 \textit{Note that this amplitude  and its associated probability distribution function
  (pdf) $f_{RFI}(x)=\psi^2_{RFI}(x)$, extremize the RFI subject to constraints.
  It is critical to note that this $I[\psi^2_{RFI}]$ does not correspond
    to the FIM scenario, that can be re-obtained by  setting $V(x)=0$ resulting in $I[\psi^2_{FIM}]$}.

A necessary condition for the derivation of (5) is that the following relation
\begin{equation}
4\left\langle {x\frac{{d\tilde U_{RFI}^{Physical} \left( x
\right)}}{{dx}}} \right\rangle_{\psi ^2_{RFI}}  = 2\left\langle {V_{xx} \left( x
\right)} \right\rangle_{\psi ^2_{RFI}}  - \left\langle {V_x^2 \left( x \right)}
\right\rangle_{\psi ^2_{RFI}}, \\
\end{equation}
be satisfied.  Here, $\tilde U_{RFI}^{Physical}(x)$ is defined by (15) in Section 3.  \textit{Appendix A provides an interpretation of (13)}.  Eq. (13) is generally satisfied in the case of many prominent QM
potentials, such as the anharmonic quartic, and sextic potentials
[24,25], when expressed in polynomial-form $
V\left( x \right) = \sum\limits_j {c_j x^j }$.

\section{Hellmann-Feynman-RFI connections}

Multiplying  (4) by 2 and re-arranging terms yields
\begin{equation}
 - \frac{{d^2 \psi_{RFI} \left( x \right)}}{{dx^2 }} + \tilde U_{RFI}
  \left( x \right)\psi_{RFI} \left( x \right) = \frac{{\lambda _0 }}{4}\psi_{RFI} \left( x \right),
\end{equation}
where the RFI pseudo-potential (14) is re-defined as
\begin{equation}
\tilde U_{RFI}(x)  = \underbrace { - \frac{1}{4}\left[ {\sum\limits_{i = 1}^M {\lambda _i A_i \left( x \right)} } \right]}_{\tilde U_{RFI}^{Data}(x)
}\underbrace { - \frac{1}{4}\left[ {2V_{xx} \left( x \right) - V_x^2 \left( x \right)} \right]}_{\tilde U_{RFI}^{Physical}(x) }.
\end{equation}
Note that (14) adopts  the form of the usual time independent Schr\"{o}dinger equation, having energy eigenvalue $E$, for $
\frac{{\hbar ^2 }}{{2m}} = 1$ and $
\frac{{\lambda _0 }}{4} = E$.
The quantum mechanical virial theorem states that [15]
\begin{equation}
-\int {\psi_{RFI} \left( x \right)\frac{{d^2 \psi_{RFI} \left( x \right)}}{{dx^2 }}} dx
= \left\langle {x\frac{{d \tilde U_{RFI}(x) }}{{dx}}} \right\rangle_{\psi ^2_{RFI}},
 \end{equation}
where $
\left\langle  \bullet  \right\rangle_{\psi ^2_{RFI}}$ denotes expectations evaluated with respect to $f_{RFI}(x)=\psi^2_{RFI}(x)$.
Following through with the derivation and invoking (13) and (15), yields
\begin{equation}
\lambda _0  + \sum\limits_{i = 1}^M {\lambda _i \left\langle {A_i \left( x \right)} \right\rangle_{\psi ^2_{RFI}} }  =  - \sum\limits_{i = 1}^M {\lambda _i
\left\langle {x\frac{{dA_i \left( x \right)}}{{dx}}} \right\rangle_{\psi ^2_{RFI}} }.
\end{equation}
Here, (17) is exctly Eq. (41) in [1].  Note that $\tilde U_{RFI}^{Physical}(x)$ is \textit{not} inferred
from the empirical observables.  \textit{Thus, even in the absence
of them, $\tilde U_{RFI}^{Physical}(x) \neq 0$ is an
a-priori specified analytical expression derived from a known
physical potential}.  Further, the a-priori specified QM potential $V(x)$ is taken to be generic in this Section.  Since the $A_i(x)$ can always be expressed
as power-series, one can write, without loss of generality invoking (15),
\begin{equation}
\tilde U_{RFI}^{Data}(x)=-\frac{1}{4}\sum\limits_k {\lambda _k A_k \left( x \right)}  = \sum\limits_k {a_k x^k } ;a_k  = -\frac{{\lambda _k }}{4}.
\end{equation}
The practical utility of (18), modified in Section 4 below, will be seen in Section 5,
 wherein the coefficients of the QM potentials obtained via inference from experimental measurements
 are seamlessly related to the concomitant Lagrange multipliers.
 Setting for simplicity $A_k(x)=x^k$, and substituting into
(17), one finds
\begin{equation}
\lambda _0  =  - \sum\limits_{k=1 }^M {\left( {k + 1} \right)\lambda_k\left\langle {A_k(x) } \right\rangle_{\psi ^2_{RFI}} .}
\end{equation}
\textit{As is demonstrated below in Section 5, (19) proves invaluable in inferring first the values
 of $\lambda_0$, and then the energy-eigenvalue $E$,
without recourse to solving (4) or (14).
It is readily verifiable that (19) is a consequence of (5),
 subjected to the reciprocity relation (8)}.  Substituting now (19) into (12) yields
\begin{equation}
\Im[\psi^2_{RFI}(x)|e^{-V(x)}] =
 - \sum\limits_{k=1 }^M {k\lambda _k \left\langle {A_k(x) } \right\rangle_{\psi ^2_{RFI}} .}
\end{equation}
We have  from (14)
\begin{equation}
H_{RFI}\psi_{RFI} \left( x \right) = \frac{{\lambda _0 }}{4}\psi_{RFI} \left( x \right) = E\psi_{RFI} \left( x \right),
\end{equation}
where the information-theoretic Hamiltonian is defined by 
$H_{RFI} = \left[ { - \frac{{d^2 }}{{dx^2 }} + \tilde U_{RFI}(x) } \right]$,
 and the energy eigenvalue relates to the normalization Lagrange multiplier as
$
E = \frac{{\lambda _0 }}{4}$.  Application of the HFT  to (21) yields the reciprocity relation (8)
\begin{equation}
\frac{\partial }{{\partial \lambda _k }}\left( {\frac{{\lambda _0 }}{4}} \right) = \left\langle \psi_{RFI}  \right| - \frac{{A_k(x){}}}{4}\left| \psi_{RFI}
\right\rangle  \Rightarrow \frac{{\partial \lambda _0 }}{{\partial \lambda _k }} =  - \left\langle {A_k(x) } \right\rangle_{\psi ^2_{RFI}}.
\end{equation}
Now, operating on (19) leads to
\begin{equation}
\frac{{\partial \lambda _0 }}{{\partial \lambda _k }} =  - \left( {j + 1} \right)\left\langle {A_k(x) } \right\rangle  - \left( {k + 1}
\right)\sum\limits_{k = 1}^M {\lambda _k } \frac{{\partial \left\langle {A_k(x) } \right\rangle_{\psi ^2_{RFI}} }}{{\partial \lambda _j }}
\end{equation}
Eqs. (22) and (23) together produce
\begin{equation}
j\left\langle {A_j(x) } \right\rangle_{\psi ^2_{RFI}}  =
  - \left( {k + 1} \right)\sum\limits_{k = 1}^M {\lambda _k }
   \frac{{\partial \left\langle {A_k(x) } \right\rangle_{\psi ^2_{RFI}} }}{{\partial \lambda _j }},
\end{equation}
and taking derivatives of (20) one encounters
\begin{equation}
\frac{{\partial \Im \left[ {\psi ^2_{RFI} \left( x \right)\left| {e^{ - V\left( x \right)} } \right.} \right]}}{{\partial \lambda _j }}=  - j\left\langle {A_j(x) } \right\rangle_{\psi ^2_{RFI}}  - k\sum\limits_{k =
 1}^M {\lambda _k } \frac{{\partial \left\langle {A_k(x) } \right\rangle_{\psi ^2_{RFI}} }}{{\partial \lambda _j }}.
\end{equation}
Finally, substituting (24) into (25) yields the generalized RFI-Euler theorem (10)
\begin{equation}
\frac{{\partial \Im \left[ {\psi ^2_{RFI} \left( x \right)\left| {e^{ - V\left( x \right)} } \right.} \right]}}{{\partial \lambda _j }} = \sum\limits_{k = 1}^M {\lambda _k
\frac{{\partial \left\langle {A_k(x) } \right\rangle_{\psi ^2_{RFI}} }}{{\partial \lambda _j }}}.
\end{equation}
On the other hand, taking derivatives of (7) with respect to $\lambda_j$,
 and comparing the ensuing result with (25), immediately leads to the reciprocity relation (9)
\begin{equation}
\frac{{\partial \Im \left( {\left\langle {A_1(x) } \right\rangle_{\psi ^2_{RFI}} ,...,\left\langle {A_M(x) } \right\rangle_{\psi ^2_{RFI}} } \right)}}{{\partial \left\langle {A_j(x) }
\right\rangle }_{\psi ^2_{RFI}}} = \lambda _j.
\end{equation}
\textit{Thus, the reciprocity relations (8) and (9) and the generalized RFI-Euler theorem (10) are shown to be self-consistently derived from the HFT}. An important results follow from substituting (27) into (20). One arrives at a linear,  partial differential
equation (PDE) for the RFI
\begin{equation}
\Im  =  - \sum\limits_{k = 1}^M {k\left\langle {A_k(x) } \right\rangle_{\psi ^2_{RFI}} \frac{{\partial \Im }}{{\partial \left\langle {A_k(x) } \right\rangle_{\psi ^2_{RFI}} }}.}
\end{equation}
Here, (28) is a qualitative extension of the result obtained in [25].
Investigations into the physical implications of (28) is the task of future
works.  The solution of (28) is
\begin{equation}
\Im \left( {\left\langle {A_1(x) } \right\rangle_{\psi ^2_{RFI}} ,...,,\left\langle {A_M(x) } \right\rangle_{\psi ^2_{RFI}} } \right) =   \sum\limits_{k = 1}^M {C_k  {\left\langle
{A_k(x) } \right\rangle } ^{\frac{{ - 1}}{k}} ,}
\end{equation}
where $C_k >0$ is a constant of integration.  Note that $
\left\langle {A_k \left( x \right)} \right\rangle _{\psi _{RFI}^2 }  = \left\langle {x^k } \right\rangle _{\psi _{RFI}^2 } $.

\section{Ansatz}
 Multiplying the LHS of (14) by four yields the FIM in amplitude form,
  subjected to a single integration-by-parts.   Thus, redefining (14) in terms of the
 pdf $f_{RFI}(x)=\psi^2_{RFI}(x)$ one finds
\begin{equation}
\int {f_{RFI}\left( x \right)\left( {\frac{{d\ln f_{RFI}\left( x \right)}}{{dx}}} \right)^2 dx}  = 4\int {f_{RFI}\left( x \right)\left( {x\frac{{d\tilde U_{RFI}(x)
}}{{dx}}} \right)} dx.
\end{equation}
Eqs. (30)and  (15) together yield
\begin{equation}
\int {f_{RFI}\left( x \right)\left[ {\left( {\frac{{d\ln f_{RFI}\left( x \right)}}{{dx}}} \right)^2
 - 4x\frac{{d\tilde U_{RFI}^{Data} }}{{dx}} - 4x\frac{{d\tilde U_{RFI}^{Empirical} }}{{dx}}} \right]dx}  = 0.
\end{equation}

In another vein, (15) and (18) are expeditiously cast to give  the empirical pseudo-potential contributions in (31)
 a useful  form
\begin{equation}
4x\frac{{d\tilde U_{REI}^{Data}(x) }}{{dx}} =  - \sum\limits_{k } {k\lambda _{k} x^{k} },
\end{equation}
while the contributions of the physical pseudo-potential in (31) are obtained from (15) in the fashion
\begin{equation}
4x\frac{{d\tilde U_{RFI}^{Physical} }(x)}{{dx}} = x\frac{{dV_x^2 \left( x \right)}}{{dx}} - 2x\frac{{dV_{xx} \left( x \right)}}{{dx}}.
\end{equation}
Now, substituting (32) and (33) into (31) yields
\begin{equation}
\int {f_{RFI}\left( x \right)\left[ {\left( {\frac{{d\ln f_{RFI}\left( x \right)}}{{dx}}} \right)^2  + \sum\limits_{k } {k\lambda _{k} x^{k} }   - x\frac{{dV_x^2 \left( x \right)}}{{dx}} + 2x\frac{{dV_{xx} \left( x \right)}}{{dx}}} \right]dx}  = 0.
\end{equation}
In (34), $f_{RFI}(x)=\psi^2_{RFI}(x)$ where $\psi_{RFI}(x)$ is obtained via solution of (4).  Eq. (4) is most generally satisfied by Hermite-Gauss polynomials, which are orthogonal to the Gaussian (exponential) distribution.  To study the leading-order contributions, specializing the solution of (34) to exponential forms requires that the terms in $[\bullet]$ satisfy
\begin{equation}
f_{RFI}\left( x \right) = N\exp \left\{ { \pm \int {\sqrt { - \sum\limits_{k } {k\lambda _{k} x^{k}  + x\frac{{dV_x^2 \left( x \right)}}{{dx}} - 2x\frac{{dV_{xx} \left( x \right)}}{{dx}}} } } } \right\}.
\end{equation}
The arbitraty constant of integration in (35) is absorbed into the normalization factor:$
N = \int\limits_{ - \infty }^\infty  {f\left( x \right)dx}  = 1$, which ensures that the pdf valishes at $\pm \infty$.  Note that the normalization factor may also be construed as being the \textit{partition function}.  Note that the more general solution for $f(x)$ in the form of Hermite-Gauss polynomials is the task of future work (see Section 7).  This is recast with the aid of (18) and  (35)  as
\begin{equation}
f_{RFI}\left( x \right) = N\exp \left\{ { \pm \int {\sqrt { \sum\limits_{k } {4ka _{k} x^{k}  + x\frac{{dV_x^2 \left( x \right)}}{{dx}} - 2x\frac{{dV_{xx} \left( x \right)}}{{dx}}} } } } \right\},
\end{equation}
where $a_{k}$ are the coefficients of the QM physical potential,
 inferred through empirical observations.
In accordance with prior studies (eq. [24, 25]) which describe commonly encountered QM potentials such as the anharmonic
quartic, and sextic potentials in terms of polynomials, the a-priori specified physical QM potential in (33)-(36) be expressed as the particular solution 
\begin{equation}
V\left( x \right) = \sum\limits_j {c_j x^j },
\end{equation}
Here, $c_j$ is a known constant given in the form of the a-priori specified physical QM potential.
The sign in the exponential is chosen, contingent to i) the nature of $V(x)$ (together with its derivative terms)
 and ii) the QM potential, in  such a manner that
    $f_{RFI}\left( x \right)$ as $x \rightarrow \pm \infty  $ is physical. Note that in (35) taking the negative value in the exponential is tenable since $
\left( {\frac{{d\ln f_{RFI}\left( x \right)}}{{dx}}} \right)^2  = \left( {\frac{{d\left( { - \ln f_{RFI}\left( x \right)} \right)}}{{dx}}} \right)^2$.
     For the sake of comparisons of the pdf profiles between the RFI and FIM scenarios, it is important
     to specify the Fisher-equivalent of (32) [18,19] by setting $V(x)=0$. This produces
the following ansatz for the pdf
\begin{equation}
\begin{array}{l}
f_{FIM}\left( x \right) = Nexp\left\{ { \pm \int {\sqrt {-\sum\limits_{k } { k\lambda _{k}^{FIM} x^{k}  } }
} dx} \right\}=Nexp\left\{ { \pm \int {\sqrt {\sum\limits_{k } { 4ka _{k} x^{k}  } }
} dx} \right\}.
\end{array}
\end{equation}
\section{Analysis and Numerical Simulations}

The theoretical results in the previous Sections
 are now employed to establish the qualitative distinction between RFI-based variational extremizations
  vis-\'{a}-vis equivalent Fisher-based ones.  This is accomplished via the ansatz derived
  in Section 4, specialized to pseudo-potentials corresponding to QM potentials.  One of the most salient results of this paper is to demonstrate the inference of the
   energy-eigenvalues of the S-RFI link,
    without recourse to solving (4).
     The treatment for this procedure is provided in this Section wherein  the necessary mathematical and
procedural "machinery" to accomplish the above task are established, commencing with a candidate physical
QM potential of the form [23]
 \begin{equation}
\tilde V^{Inf}\left( x \right) = \omega^2 x^2+ \varepsilon x^{k};~k=4,6,...,
\end{equation}
inferred from the empirical observables $
\left\langle {x^{k} } \right\rangle ;k = 2,4,...$.
Here, $\varepsilon$ is the anharmonicity constant.  Generically, (39) may be cast in the form
\begin{equation}
\tilde V^{Inf} \left( x \right) = \sum\limits_{k} {a_{k} x^{k} };k=2,4,...,
\end{equation}
where $a_{k}$ are coefficients  related to the concomitant Lagrange multipliers through (18).

\textit{An observation of much significance arises here: while the energy-eigenvalues and the concomitant pdf profiles
pertaining to the RFI
 framework noticeably differ from their FIM counterparts, the Lagrange multipliers for the constraint
 terms, corresponding to the experimental observables, are readily  demonstrated to be the
 same for both the RFI and the FIM scenarios}. The cause for this somewhat
 surprising finding is traced to the \textit{Gibbs form}  utilized in (2) and (3) of this paper.
 Specifically, on account of the \textit{Gibbs form} in (2) and (3),
 the S-RFI link (4), (14), and (15) have the Lagrange multipliers
 corresponding to the inference process completely separated and delineated from the a-priori specified knowledge
 denoted by the derivatives of the physical QM potential $V(x)$.  Note that within the RFI framework, the empirical observables are explicitly defined as $
\left\langle {x^k } \right\rangle _{\psi _{RFI}^2 }$.  Likewise, in the FIM-case the empirical observables are defined as $
\left\langle {x^k } \right\rangle _{\psi _{FIM}^2 }$.

This Section treats the cases of both perturbed and non-perturbed
QM potentials, inferred from empirical observables, viz. the harmonic oscillator (HO, hereafter)
potential and the quartic anharmonic oscillator (QAHO, hereafter) potential, respectively.
For all cases herein, the a-priori specified physical QM potential is taken to be the simple HO potential
\begin{equation}
V\left( x \right) = \omega^2x^2.
\end{equation}
In this paper, all computations are performed using $MATHEMATICA^{\circledR}$.
\subsection{Procedure for the inference process}
$(i)$~~Given empirical observables $
\left\langle {x^k } \right\rangle _{\psi _{RFI,FIM}^2 }$, a candidate (inferred) QM potential of the form given by (40) is specified.  \textit{Note that the pdf expressed in the form of amplitudes $\psi _{RFI}^2$ and $\psi _{FIM}^2$ which defines the expectations of the empirical observables depends solely upon whether the RFI or the FIM frameworks are being studied, and does not influence the choice of the inferred QM potential}.  \\
$(ii)$~~Invoking (14) and (15), yields the S-RFI link
\begin{equation}
\left[ { - \frac{{d^2 }}{{dx^2 }} - \sum\limits_{k } {a_{k}x^{k} }  + \tilde U_{RFI}^{Physical}(x) } \right]\psi_{RFI} \left( x \right) = E\psi_{RFI} \left( x \right).
\end{equation}
$(iii)$~~From (18), the Lagrange multipliers $\lambda_{k}$ are related to the coefficients $a_{k}$ via (40).\\
$(iv)$~~The physical contributions in (35) are obtained by seeking consistency between (37) and (41), which requires
\begin{equation}
c_2  = \omega ^2 ;V_x^2 \left( x \right) = 4\omega ^4 x^2 ;2V_{xx} \left( x \right) = 4\omega ^2 ;j= 2.
\end{equation}
Thus,
\begin{equation}
x\frac{{dV_x^2 \left( x \right)}}{{dx}} = 8\omega ^4 x^2,  {\rm and,} 2x\frac{{dV_{xx} \left( x \right)}}{{dx}} = 0.
\end{equation}
$(v)$~~The inferred pdf is then  obtained from (35), which are
employed to obtain the moments in (19) thereby the $\lambda_0$
value. From Section 3, the energy-eigenvalue is $E =
\frac{{\lambda _0 }}{4}$. \textit{Thus, inference of the
energy-eigenvalues is accomplished without solving the
Schr\"{o}dinger-like
link for the RFI}. \\
$(vi)$~~To benchmark the inference process with published results expressed either in terms of the Schr\"{o}dinger wave equation (SWE,
 hereafter) [23] or its information theoretic counterpart derived in FIM-based studies [22],
 it is imperative that $V(x)=0$ be specified in the RFI results presented in this paper.
  This is accomplished for the case of the HO and the QAHO potentials in Section 5.2.\\
$(vii)$~~Setting $V(x)=0$ in (42) yields the Schr\"{o}dinger-like equation for the FIM-framework.
\begin{equation}
\left[ { - \frac{{d^2 }}{{dx^2 }} - \sum\limits_{k } {a_{k}x^{k} } } \right]\psi_{FIM} \left( x \right) = E_{FIM} \psi_{FIM} \left( x \right).
\end{equation}

\subsection{Test cases}
\subsubsection{Non-perturbed case~$\varepsilon =0$}
Let the empirical measurements be $
\left\langle {x^2 } \right\rangle _{\psi _{RFI,FIM}^2 }$, depending upon whether the RFI or the FIM frameworks are being studied.  Thus, an empirically inferred QM  candidate  potential is
the HO of the form
\begin{equation}
\tilde V^{Inf}\left( x \right) = \omega^2x^2.
\end{equation}
The a-priori specified QM physical potential is also taken to be of the form described by (41).
Note that the distinction between the RFI and FIM frameworks is
obvious,  even in the simple cases of the HO
solution.

Invoking (18) yields
\begin{equation}
\lambda_2=-4\omega^2.
\end{equation}
The inferred pdf is thus is thus obtained from (35) with the aid of (44) and (47) as
\begin{equation}
f_{RFI}\left( x \right) = N\exp \left\{ { - \int {\sqrt {8\omega ^2 x^2(1+\omega^2) } } dx} \right\}=
N\exp \left\{ { - \sqrt {\left( {2 \omega ^2 \left( {1 + \omega ^2 } \right)} \right)} x^2 } \right\},
\end{equation}
where $N$ is a normalization factor.  Setting $\omega^2=0.5$ yields
\begin{equation}
\int\limits_{ - \infty }^\infty  {f\left( x \right)dx}  = 1 \Rightarrow
N = 0.624378.
\end{equation}

For $k=2$ and $\omega^2=0.5$, the normalization Lagrange multiplier is explicitly obtained from (19),
with the aid of (47) and (48),  as
$ \lambda_0
 \approx 2.44596$.  From Section 3, $E = \frac{{\lambda _0 }}{4} \approx  0.611489$.
\textit{Note that, from (42) and (15), it is evident that $E$ is not the energy-eigenvalue for the harmonic oscillator potential.
 Specifically, it corresponds instead to a different potential, namely,  a
superposition of the empirical HO contributions and the derivative terms of the a-priori
specified QM potential (41), which in this study is also the HO potential}.

To benchmark the inference process, comparison between the second term on the LHS of (45) with (46) clearly shows
$a_{k}=\omega^2$ for $k=2$.  Invoking (18) yields
\begin{equation}
\lambda^{FIM}_2=-4\omega^2.
\end{equation}
This is identical to (47).  Carrying through with the analysis in
a manner exactly similar to that given above, the inferred pdf
from the FIM framework is obtained with the aid of (38) for
$\omega^2=0.5$  as
\begin{equation}
f_{FIM}\left( x \right) =
N\exp \left\{ -x^2 \right\},
\end{equation}
which exhibits the correct HO-form, where the normalization factor $N$ is
\begin{equation}
\int\limits_{ - \infty }^\infty  {f_{FIM}\left( x \right)dx}  = 1 \Rightarrow
N =
\frac{1}{{\sqrt \pi  }},
\end{equation}
which is the exact theoretical result as given in [22].

 Along the lines of [22],  the inferred energy-eigenvalue is theoretically obtained by specifying
\begin{equation}
\left\langle {x^2 } \right\rangle_{\psi^2_{FIM}}  = \frac{9}{{64\omega ^4 }}.
\end{equation}
Setting $V(x)=0$, $k=2$, and $C_2=1$ in (29) yields
\begin{equation}
I\left[ {\psi^2 _{FIM} } \right] = \left\langle {x^2 } \right\rangle ^{ - \frac{1}{2}}_{\psi^2_{FIM}}.
\end{equation}
Setting $V(x)=0$ in the reciprocity relation (9) and substituting (54) into the resulting expression gives for $k=2$
\begin{equation}
\lambda _2^{FIM}  =  - \frac{1}{2}\left\langle {x^2 } \right\rangle ^{ - \frac{3}{2}}_{\psi^2_{FIM}}  =  - 4\omega ^2.
\end{equation}
Setting $V(x)=0$ and $k=2$ into (12), the FIM-case is obtained as
\begin{equation}
\lambda _0^{FIM}  = I\left[ {\psi^2_{FIM} } \right] - \lambda _2^{FIM} \left\langle {x^2 } \right\rangle_{\psi^2_{FIM}}=
\frac{3}{2}\left\langle {x^2 } \right\rangle ^{ - \frac{1}{2}}_{\psi^2_{FIM}}.
\end{equation}
Substituting (54) and (55) into (56), invoking (53), and setting
$\omega^2=0.5$ into the resultant yields $\lambda^{FIM}_0=2.0$.
From Section 3, the inferred energy-eigenvalue $ E = \frac{1}{2}$
is the desired result.  Note that unlike the analysis in [22],
(54) does not saturate the Cramer-Rao bound [27].  This issue will
be discussed in Section 5.3 (below).
\subsubsection{Perturbed case~$\varepsilon \ne 0$}
Let the empirical measurements be $
\left\langle {x^2 } \right\rangle _{\psi _{RFI,FIM}^2 }$ and $
\left\langle {x^4 } \right\rangle _{\psi _{RFI,FIM}^2 }$.  Thus, an  empirically inferred QM candidate potential is
the QAHO of the form [23]
\begin{equation}
\tilde V^{Inf}\left( x \right) = \omega^2x^2 + \epsilon x^4,
\end{equation}
where $\epsilon$ is the anharmonicity constant.  The a-priori specified QM physical potential
is again taken to be of the form described by (40).  \textit{Note that the pdf expressed in the form of amplitudes $\psi _{RFI}^2$ and $\psi _{FIM}^2$ which defines the expectations of the empirical observables depends solely upon whether the RFI or the FIM frameworks are being studied, and does not influence the choice of the inferred QM potential (57)}.
  Comparison between the second term on the LHS of (42) with (57) clearly shows $a_{2}=\omega^2$
  together with  $a_{4}=\varepsilon$.  Invoking now (18) yields
\begin{equation}
\lambda_2=-4\omega^2, {\rm and,} \lambda_4=-4\varepsilon.
\end{equation}
Since the a-priori specified QM potential is the simple HO,
the physical portion of the ansatz for the inferred pdf (35) is identical to (43) and (44).
 The inferred pdf is thus is thus obtained from (37) with the aid of (44) and (58) as
\begin{equation}
\begin{array}{l}
f_{RFI}\left( x \right) = N\exp \left\{ { - \int {\sqrt {8\omega ^2(1+2\omega^2)x^2 +16\varepsilon x^4} } dx} \right\}\\
\\
=N\exp \left\{ { -  \frac{\sqrt 2{\left[ {x^2 \left( {\omega ^2  + 2\omega ^4  + 2\varepsilon x^2 } \right)} \right]^{\frac{3}{2}} }}{{3\varepsilon x^3 }}}  \right\}=
N\exp \left\{ { - \frac{{\sqrt 2 }}{{3\varepsilon }}\left[ {\omega ^2  + 2\omega ^4  + 2\varepsilon  x^2 } \right]^{\frac{3}{2}} } \right\},
\end{array}
\end{equation}
where $N$ is a normalization factor.

This paper seeks to benchmark the inferred energy-eigenvalues of the QAHO with established numerical solutions for the SWE within the limit $V(x)=0$.  Ref. [23] employs the following scaling relation as its basis
\begin{equation}
a^2 E_n^k \left( {\omega_{SWE}^2 ,\varepsilon_{SWE} } \right) \approx E_n^k \left( {\omega_{SWE}^2 a^4 ,\varepsilon_{SWE}a^{k + 2} } \right),
\end{equation}
where $n$ is the quantum number (which is zero for this study), and $a>0$.  For the QAHO, $k=4$.  Thus, for $\omega_{SWE}^2=\varepsilon_{SWE}=1$, let $a^2=0.5$, and thus $\omega^2=\omega_{SWE}^2 a^4=0.25$ and $\varepsilon=\varepsilon_{SWE} a^6=0.125$.

From (19) for $k=2,4$, the normalization Lagrange multiplier is explicitly stated in terms of its moments as
\begin{equation}
\lambda _0  = 3.0\left\langle {x^2 } \right\rangle_{\psi^2_{RFI}}  + 2.5\left\langle {x^4 } \right\rangle_{\psi^2_{RFI}} .
\end{equation}
Owing to the highly oscillatory nature of the integrals over the entire range $[-\infty,\infty]$, convergence cannot be guaranteed in general.  Thus, the moments in (61) are numerically evaluated in $[-2.0,2.0]$ yielding $N=1.35953$, $\lambda_0 \approx 2.54677$ and $E \approx 0.636693$.
\textit{Again, remark that, from (42) and (15), $E$ is not the  QMHO potential energy-eigenvalue,
but that for a potential
which is a superposition of the HO contributions and the a-priori specified QM potential (41)
}.

For benchmarking purposes, we carry through with an analysis analogous  to the that for the  HO-case in Section 5.2.1.
 Accordingly, setting $V(x)=0$ yields
\begin{equation}
\lambda^{FIM}_2=-4\omega^2,and,\lambda^{FIM}_4=-4\varepsilon.
\end{equation}
The inferred pdf for the case of the quartic anharmonic oscillator
in the FIM framework obtained with the aid of (38) is thus
\begin{equation}
f_{FIM}\left( x \right) = N\exp \left\{ { - \frac{{\sqrt 2 }}{{3\varepsilon }}\left[ {\omega ^2  + 2\varepsilon  x^2 } \right]^{\frac{3}{2}} } \right\},
\end{equation}
where $N$ is a normalization factor.

For the case of $\omega^2=0.25,\varepsilon=0.125$, the energy-eigenvalue obtained from (19),(60)-(63) is $E \approx 0.72176$.
The energy-eigenvalue from [23] scaled by $a^2=0.5$ for the quantum number $n=0$ and $\omega^2=1.0,\varepsilon=1.0$ is $E^{SWE} \approx 0.696176$.  Note that the equivalent expressions for the QAHO case in the FIM-study [22] yield a value of the energy-eigenvalue $E \approx 0.320024$ for $\omega^2=0.25,\varepsilon=0.125$, which represents a highly inaccurate inference.  Further, the case of $\omega^2=0.5,\varepsilon=0.353553$ yields the energy-eigenvalue obtained from (19), (60)-(63) as $E \approx 1.0936$. The corresponding energy-eigenvalue from [23] scaled by $a^2=0.707107$ for the quantum number $n=0$ and $\omega^2=1.0,\varepsilon=1.0$ is  $E^{SWE} \approx 0.984541$.
 The QAHO energy-eigenvalue obtained from the inference model in this paper
 displays an excellent degree of consistency with the scaled energy-eigenvalues for the SWE [23], for a wide range of coefficient values of the inferred QM potential.

\subsection{Comments and numerical simulations}

    The peculiar circumstance that makes identical  the Lagrange multipliers for both the RFI and the FIM frameworks
is a consequence of the \textit{Gibbs form} adopted in (2) and (3).
In Section 5.2.1, the energy-eigenvalue for the HO potential inferred from empirical observables shows excellent consistency with theoretical results within the limit $V(x)=0$ [22,23].  This precision of the inference process described in this paper carries over to the case of the QAHO potential in Section 5.2.2 within the limit $V(x)=0$, when compared with scaled energy-eigenvalues that are numerically obtained [23].  Figs. 1 and 2 depict the inferred pdf's for the RFI and FIM frameworks for the HO and the QAHO, respectively.   In both cases, it is readily observed that the inferred pdf's obtained from the RFI framework are more sharply peaked than their FIM and SWE counterparts.

It is important to note that for expressions in this paper to reduce to those in
 prior FIM-studies [18,19,22,26] in the limit $V(x)=0$, the constant in (19) and (20) $
k \to \frac{k}{2}$.  This discrepancy arises on account of the manner
in which the empirical pseudo-potential $\tilde U^{Data}_{RFI}(x)$ and the physical
 pseudo-potential $\tilde U^{Physical}_{RFI}(x)$ are defined in (14) and (15).
   Likewise, for (29) to reduce to its FIM-counterpart as defined in previous studies (eg. [20,
   22]),  in the limit $V(x)=0$, the power in (29) in this paper becomes $
 - \frac{1}{k} \to  - \frac{2}{k}$.  It is for this reason that the FIM in (54) does
  not saturate the Cramer-Rao bound [27] in a manner similar to its counterpart in [22].
  Additionally, this discrepancy (in the limit $V(x)=0$) between the expressions in this paper
  and those in previous FIM studies results in differences in the values of the
   normalization Lagrange multiplier and, consequently, the inferred energy-eigenvalues.

\section{Reconstruction of the RFI-framework }

This Section establishes the theoretical basis to reconstruct the RFI-framework from the FIM-case given: $(i)$  the
values of the energy-eigenvalues of the FIM model, $(ii)$  the
empirical observations $<x^k>_{\psi^2_{FIM}}$ (and thus the coefficients of the
inferred QM potential), and $(iii)$  the form of the a-priori
specified QM potential.  This is accomplished with the aid of (13)
and the fact that the Lagrange multipliers $\lambda_k$ associated
with the constraint terms are identical for both the RFI and the
FIM frameworks, a peculiarity which is traced to the {Gibbs form}
employed in (2) and (3).  Note that, owing to the transitions
between the FIM and the RFI frameworks in this Section, all
quantities (eg. Lagrange multipliers, expectation values,
amplitudes, and pdf's), are explicitly associated with the measure
of uncertainty that they are associated with.  

For example, the
expectations $<\bullet>_{\psi^2_{RFI}}$ and $<\bullet>_{\psi^2_{FIM}}$ are evaluated
with respect to $f_{RFI}(x)$ and $f_{FIM}(x)$, the pdf's which
extremize the RFI and the FIM, respectively.  With the aid of
(18), and setting $ E^{FIM}  = \frac{{\lambda _0^{FIM} }}{4}$, one
has from (45)
\begin{equation}
 - \frac{{d^2 \psi _{FIM} \left( x \right)}}{{dx^2 }} - \frac{1}{4}\sum\limits_k {\lambda _k A_k \left( x \right)} \psi _{FIM} \left( x \right) = \frac{{\lambda _0^{FIM} }}{4}\psi _{FIM} \left( x \right).
\end{equation}
Multiplying (64) by $4\psi _{FIM} \left( x \right)$ and integrating yields, on
invoking the virial theorem (30) and setting $V(x)=0$,
\begin{equation}
I\left[ {\psi _{FIM}^2 } \right] = \lambda _0^{FIM}  + \sum\limits_k {\lambda _k \left\langle {x_k } \right\rangle _{\psi _{FIM}^2 } } .
\end{equation}
Note that (65) has taken advantage of the fact discussed in Section 5, that the \textit{Gibbs form}
employed in (2) and (3) permits setting
$\lambda^{RFI}_k=\lambda^{FIM}_k=\lambda_k$.   Expressing (65) in terms of pdf's, leads to
\begin{equation}
\int {f_{FIM} \left( x \right)\left[ {\left( {\frac{{d\ln f_{FIM} \left( x \right)}}{{dx}}} \right)^2  - \lambda _0^{FIM}  - \sum\limits_k {\lambda _k x^k } } \right]dx = 0}.
\end{equation}
Eq. (66) is satisfied by specifying $f_{FIM} \left( x \right)$ as an exponential form resulting in
\begin{equation}
f_{FIM} \left( x \right) = \exp \left\{ { - \int {\sqrt {\lambda _0^{FIM}  + \sum\limits_k {\lambda _k x^k } } } dx} \right\}.
\end{equation}
Invoking (17) for the FIM-framework, it is readily observed that (67) is a
manifestation of (38). Comparison of (66) and (34) for $V(x)=0$ yields
\begin{equation}
\sum\limits_k {k\lambda _k x^k }  =  - \lambda _0^{FIM}  - \sum\limits_k {\lambda _k x^k }.
\end{equation}
Note that even in the case $V(x) \ne 0$, the term  $\sum\limits_k {k\lambda _k x^k }$ corresponds to the empirical pseudopotential in (15) and is independent of the a-priori specified QM potential.  Specifying in (66)
\begin{equation}
\lambda _0^{RFI}  = \lambda _0^{FIM}  + x\frac{{dV_x^2 \left( x \right)}}{{dx}} - 2x\frac{{dV_{xx} \left( x \right)}}{{dx}},
\end{equation}
results in the transition from $f_{FIM}(x) \rightarrow f_{RFI}(x)$ and subsequently the transition from the $
FIM \to RFI$ frameworks, yielding
\begin{equation}
\int {f_{RFI} \left( x \right)} \left[ {\left( {\frac{{d\ln f_{RFI} \left( x \right)}}{{dx}}} \right)^2  - \lambda _0^{RFI}  - \sum\limits_k {\lambda _k x^k } } \right] = 0
\end{equation}
For exponential forms of $f_{RFI}(x)$, the solution of (70) is
\begin{equation}
f_{RFI} \left( x \right) = \exp \left\{ { - \int {\sqrt {\lambda _0^{RFI}  + \sum\limits_k {\lambda _k x^k } } } dx} \right\}.
\end{equation}
Multiplying (69) by $f_{RFI}(x)$ and integrating, and, invoking (33) and (13), results in
\begin{equation}
\begin{array}{l}
 \lambda _0^{FIM}  = \lambda _0^{RFI}  - \left\langle {x\frac{{dV_x^2 \left( x \right)}}{{dx}}} \right\rangle _{{\psi^2_{RFI}}}  + 2\left\langle {x\frac{{dV_{xx} \left( x \right)}}{{dx}}} \right\rangle _{{\psi^2_{RFI}}}  \\
\\
 \mathop  = \limits^{\left( {33} \right)} \lambda _0^{RFI}  - 4\left\langle {\frac{{d\tilde U_{RFI}^{Physical}(x) }}{{dx}}} \right\rangle _{{\psi^2_{RFI}}} \mathop  = \limits^{\left( {13} \right)} \lambda _0^{RFI}  - 2\left\langle {V_{xx} \left( x \right)} \right\rangle _{{\psi^2_{RFI}}}  + \left\langle {V_x^2 \left( x \right)} \right\rangle _{{\psi^2_{RFI}}}.  \\
 \end{array}
\end{equation}
Similarly, multiplying (69) by $f_{FIM}(x)$ and integrating, and, invoking (33) and (13), results in
\begin{equation}
\begin{array}{l}
 \lambda _0^{RFI}  = \lambda _0^{FIM}  + \left\langle {x\frac{{dV_x^2 \left( x \right)}}{{dx}}} \right\rangle _{{\psi^2_{FIM}}}  - 2\left\langle {x\frac{{dV_{xx} \left( x \right)}}{{dx}}} \right\rangle _{{\psi^2_{FIM}}}  \\
\\
 \mathop  = \limits^{\left( {33} \right)} \lambda _0^{FIM}  + 4\left\langle {\frac{{d\tilde U_{RFI}^{Physical}(x) }}{{dx}}} \right\rangle _{{\psi^2_{FIM}}} \mathop  = \limits^{\left( {13} \right)} \lambda _0^{FIM}  + 2\left\langle {V_{xx} \left( x \right)} \right\rangle _{{\psi^2_{FIM}}}  - \left\langle {V_x^2 \left( x \right)} \right\rangle _{{\psi^2_{FIM}}}.  \\
 \end{array}
\end{equation}

Simultaneously subtracting and adding (13) from (65) yields
\begin{equation}
\begin{array}{l}
 I\left[ \psi^2  \right] - 2\left\langle {V_{xx} \left( x \right)} \right\rangle_{\psi^2}  + \left\langle {V_x^2 \left( x \right)} \right\rangle_{\psi^2}  \\
  \\
  + \lambda _0^{FIM}  + 2\left\langle {V_{xx} \left( x \right)} \right\rangle_{\psi^2}  - \left\langle {V_x^2 \left( x \right)} \right\rangle_{\psi^2}  + \sum\limits_k {\lambda _k \left\langle {x^k } \right\rangle }_{\psi^2}  = 0. \\
 \end{array}
\end{equation}
Note that in (74), the nature of the probability with which the
expectation in the term $ \sum\limits_k {\lambda _k \left\langle
{x^k } \right\rangle  }_{\psi^2}$ is evaluated is \textit{deliberately unspecified}.  Further, it is
\textit{deliberately unspecified} as to whether $\psi$ extremizes the RFI
or the FIM. From (11), the first three terms in the LHS of (74) constitute the RFI $
\Im \left[ {\left. {\psi _{RFI}^2 } \right|e^{ - V\left( x \right)} } \right]$ if $I[\psi^2]=I[\psi^2_{RFI}]$ and $<\bullet>=<\bullet>_{{{\psi^2_{RFI}}}}$.  From (72), it is evident that next three
terms in the LHS of (74) comprise $\lambda^{RFI}_0$, and the expectations are specified as: $\left\langle {V_{xx} \left( x \right)} \right\rangle_{\psi^2 _{RFI}}$ and  $\left\langle {V_x^2 \left( x \right)} \right\rangle_{\psi^2_{RFI}}$.  Specifying $ \sum\limits_k
{\lambda _k \left\langle {x^k } \right\rangle  }= \sum\limits_k
{\lambda _k \left\langle {x^k } \right\rangle_{\psi^2_{RFI}}}$, (12) is recovered.  \textit{The introduction of the expectations of the derivatives of the physical potential via (72) facilitates the transition from the $FIM \rightarrow RFI$ frameworks.   Note that (12) constitutes one of the most fundamental relations governing the RFI framework.  Thus, (65) in conjunction with (11) and (72) allows for the reconstruction of the RFI framework from the FIM framework}.

In a similar vein, substituting (11) into (12) and suitably manipulating the pertinent terms results in
\begin{equation}
\begin{array}{l}
 \Im \left[ {\left. {\psi _{RFI}^2 } \right|e^{ - V\left( x \right)} } \right] - \left\langle {V_x^2 \left( x \right)} \right\rangle _{\psi _{RFI}^2 }  + 2\left\langle {V_{xx} \left( x \right)} \right\rangle _{\psi _{RFI}^2 }  \\ 
\\
  = \lambda _0^{RFI}  - \left\langle {V_x^2 \left( x \right)} \right\rangle _{\psi _{RFI}^2 }  + 2\left\langle {V_{xx} \left( x \right)} \right\rangle _{\psi _{RFI}^2 }  + \sum\limits_k {\lambda _k \left\langle {x^k } \right\rangle } . \\ 
 \end{array}
\end{equation}

Again, note that in (75) the nature of the probability with which the
expectation in the term $ \sum\limits_k {\lambda _k \left\langle {x^k } \right\rangle } $ is evaluated is \textit{deliberately unspecified}.
From (11), the LHS of (75) is $I[\psi^2]$, where again it is
\textit{deliberately unspecified} as to whether $\psi$ extremizes the RFI
or the FIM. However, from (72) it is evident that first three
terms in the RHS of (75) comprise $\lambda^{FIM}_0$. Specifically,
there is no explicit dependence upon the a-priori specified physical QM potential,
whose derivative terms have been absorbed into $\lambda^{FIM}_0$.
Thus, specifying $I[\psi^2]=I[\psi^2_{FIM}]$ and  $ \sum\limits_k {\lambda _k \left\langle {x^k } \right\rangle }= \sum\limits_k
{\lambda _k \left\langle {x^k } \right\rangle }_{_{\psi^2_{FIM}}}$, (65) is
recovered.  \textit{Thus, (72) in conjunction with (11) and (12)
offers an alternate to setting $V(x)=0$ in achieving a transition
from the $RFI \rightarrow FIM$ frameworks}.

The workings of the above reconstruction
 procedure are now exemplified with the aid of a simple example.
 The inference of the energy-eigenvalues with out recourse to the S-RFI link (4) and (14)
  constitutes one of the primary objectives of this paper.
  This is described in Section 5.2.
  Section 5.2.1 demonstrated excellent consistency between the
  theoretical results for the inference process presented in
   this paper with published works [22],
   for the case of the HO potential for $\omega^2=0.5$.
   This was accomplished by setting the a-priori specified physical QM potential $V(x)=0$,
   thereby converting the RFI-framework presented herein into an FIM-framework.
   The energy-eigenvalue and the normalization Lagrange multiplier for the FIM-case presented
   in Section 5.2.1 for the inferred HO potential are $E^{FIM} =0.5$ and $\lambda^{FIM}_0=2.0$, respectively.

For the case of the a-priori specified QM potential (that is again
the HO expression (41)), invoking (73) and (44) results in
\begin{equation}
\lambda _0^{RFI}  = 2.0 + 8\omega ^4 \left\langle {x^2 } \right\rangle _{{\psi^2_{FIM}}}.
\end{equation}
With the aid of (53), $\lambda _0^{RFI}=3.125$ and $E =
\frac{{\lambda _0^{RFI} }}{4} = 0.78125$. In contrast to the
\textit{approximate} inferred values of $\lambda _0^{RFI}$ and $E$
presented in Section 5.2.1 (obtained from (19), (48), and (49)),
the values presented in this Section represent an inference of the
RFI energy-eigenvalues that uses  \textit{exact} theoretical
results for the FIM-case.  Reasons  for the discrepancy between
the inferred energy-eigenvalues presented in this Section for the RFI
framework, and those of the type presented in Section 5.2.1, are the objective
of on-going work, as are obtaining high precision numerical
solutions of the SWE (instead of theoretical results for the
FIM-case) for more complex inferred QM potentials.

\section{Summary and conclusions}

The derivation of the reciprocity relations for the RFI have been
self-consistently established in this paper using the
Hellmann-Feynman theorem. The inference of an ansatz for pdf's possessing an exponential form for the RFI framework has been obtained solely on the
basis of the S-RFI link and the QM virial
theorem.  The qualitative distinctions between the RFI and the FIM
frameworks have been established.  The energy-eigenvalues of (4) have been self
consistently inferred solely on the basis of the ansatz (derived in
Section 4). This has been demonstrated in Section 5 for the case of both the harmonic oscillator and the quartic anharmonic oscillator potentials.  Such an inference has been made
possible by solely utilizing the Legendre transform structure of the RFI, and the fundamental properties of
the QM virial theorem embedded in Eq. (19) in Section 3.

Apart from the analysis and numerical simulations described in Section 5, one of the most poignant examples of the
distinction between RFI and FIM variational extremizations is stated in Section 3.  Specifically, the RFI framework yields a solution even of there are
no empirical observables.  \textit{Thus, the Quantum square well example provided in Ref. [18] cannot hold true in the RFI framework, since the
physical pseudo-potential of (4) and (14) is non-vanishing}.  Further, a most noteworthy observation, highlighted in Section 5, is that the \textit{Gibbs form} adopted in (2) and (3) renders the Lagrange multipliers associated with the empirical observables identical for both the RFI and the FIM frameworks.

There is a pronounced difference between the methodology employed in this paper to infer energy-eigenvalues and pdf's, and that employed in previous FIM studies [18, 22].  While relations analogous to the one between the Lagrange multipliers and the coefficients of the QM potential inferred from empirical observables (18) have been derived in [18, 22], Ref. [18] makes no attempt to infer the energy-eigenvalues and obtain the pdf profiles.  In contrast, this paper presents results for the inferred pdf's for \textit{both} the RFI and the FIM frameworks, and accomplishes inferring the energy-eigenvalues for the RFI framework for inferred QM potentials of \textit{both} the harmonic oscillator and the quartic anharmonic oscillator models.  Next, in the case of [22] there are pronounced distinctions between the methodology employed therein and that followed in this paper.

Specifically:  $(i)$ for the case of the harmonic oscillator (HO,
hereafter) potential inferred from empirical observables, [22]
first assumes $\left\langle {x^2 } \right\rangle _{\psi _{FIM}^2 } $ and then further \textit{specifies} that the
FIM saturates the Cramer-Rao bound.  In contrast, no such
specifications are possible in this paper because of the
differences between (19), (20), and (29) herein and their
counterparts in previous FIM-studies.  Specifically, these
differences stem from the fact that the Schr\"{o}dinger-like link
for the RFI (14) in the limit $V(x)=0$ is of the more fundamental
form of the Schr\"{o}dinger wave equation (SWE, hereafter) in
[23], instead of the "scaled" version in [22],  $(ii)$  for the
case of the quartic anharmonic oscillator (QAHO, hereafter), the
results of the inferred energy-eigenvalues in [22] cannot be
benchmarked with the more fundamental ones in [23] without
recourse to  \textit{restrictive} and ad-hoc "adjustments".  In
contrast, the procedure employed in this study prescribes no such
"adjustments" for benchmarking the inferred energy-eigenvalues for
both the RFI and the FIM frameworks. Instead, a fundamental
scaling relation (60) underlying the basis for the work presented
in [23] is employed ,  and $(iii)$ while [26] does not present
numerical results for the nature of the inferred pdf profiles,
this paper presents such cases for \textit{both} the RFI
\textit{and} the FIM frameworks, for \textit{both} inferred
harmonic oscillator \textit{and} quartic anharmonic oscillator
potentials.  Finally, it is important to note that in the QAHO
case, the energy-eigenvalues obtained via   the inference
procedure demonstrated in this paper show greater consistency with
the numerical results in [23] than those described in [22].

The work presented herein is the generalization
 of prior studies within the FIM framework [18, 19, 22, 26]
 to the case of the RFI framework. The results presented herein
  establishes the basis for a comprehensive comparison between the RFI
  and the FIM frameworks using established theoretical and/or
  numerical results for HO and QAHO models  [22,23].  Further, Sections 4 and 5 of this paper establish the qualitative
distinction between the RFI and FIM frameworks
based on an inferred ansatz of exponential form.
 Finally, Section 6 of this paper establishes the theoretical framework
 for reconstructing RFI pdf's from the FIM pdf's obtained from inference,
 employing the ansatz derived in Section 4.
 In addition to establishing the transition between the $FIM \rightarrow RFI$ frameworks,
 the analysis prescribes a method to achieve a transition from the $RFI \rightarrow FIM$
 frameworks without recourse to setting the a-priori specified potential $V(x)=0$.

The work presented in this
paper may be readily extended to more complex empirical and physical pseudo-potentials.
Specifically, such an endeavor would entail the
evaluation of a number of Lagrange multipliers.  In this case, a potentially attractive and credible  ansatz for the inferred pdf is
\begin{equation}
f_{RFI}\left( x \right) = \exp \left\{ { - \int {\sqrt { - \sum\limits_{k = 1}^M {k\lambda _{k} x^{k}  + x\frac{{dV_x^2 \left( x \right)}}{{dx}} - 2x\frac{{dV_{xx} \left( x \right)}}{{dx}}} } } } \right\}H_n(x),
\end{equation}
where $H_n(x)$ are Hermite-Gauss polynomials.  A candidate
approach to accomplish this task is the extension of the framework
presented in this paper via a suitable adaptation of the
information-theoretic optimization scheme described in [28, 29].  The game-theoretic aspects of [28, 29] have their origins in [30].
The process of inference constitutes the basis of machine learning
[31]. \vskip 4mm

 The process of obtaining the energy-eigenvalues
in Section 5 and the transitions between the $ RFI \Leftrightarrow
FIM$ frameworks in Section 6
 lend
immense credence to the prospect of formulating a comprehensive
 inference framework for compressed sensing [32].  These studies are the task of ongoing works.

\section*{Acknowledgements}

Gratitude is expressed towards the reviewers for their invaluable comments.  RCV gratefully acknowledges support from NSFC-CS $\&$ NI contract
\textit{111017-01-2013}.  
\section*{Appendix A: Interpretation of Eq. (13)}
 \renewcommand{\theequation}{A.\arabic{equation}}
  \setcounter{equation}{0}  
A necessary condition for the RFI derivations is the following
relation [Cf. Eq. (13)]:
\begin{equation}
4\left\langle {x\frac{{d\tilde U_{RFI}^{Physical} \left( x
\right)}}{{dx}}} \right\rangle_{{\psi^2_{RFI}}}  = 2\left\langle {V_{xx} \left( x
\right)} \right\rangle_{{\psi^2_{RFI}}}  - \left\langle {V_x^2 \left( x \right)}
\right\rangle_{{\psi^2_{RFI}}}. \\
\end{equation}

The RFI pseudo-potential  was re-defined in (15) as
\begin{equation}
\tilde U_{RFI}(x)  = \underbrace { - \frac{1}{4}\left[
{\sum\limits_{i = 1}^M {\lambda _i A_i \left( x \right)} }
\right]}_{\tilde U_{RFI}^{Data}(x) }\underbrace { - \frac{1}{4}\left[
{2V_{xx} \left( x \right) - V_x^2 \left( x \right)}
\right]}_{\tilde U_{RFI}^{Physical}(x) }.
\end{equation}

(A.1) and (A.2) imply that
\begin{equation}
x\frac{{d\tilde U_{RFI}^{Physical}(x) }}{{dx}} =  - \tilde U_{RFI}^{Physical}(x). 
\end{equation}

This entails
\begin{equation}
\int {x\frac{{d\tilde U_{RFI}^{Physical} \left( x \right)}}{{dx}}dx}  =  - \int {\tilde U_{RFI}^{Physical} \left( x \right)dx}. 
\end{equation}

Subjecting the LHS of (A.4) to a single integration-by-parts yields

\begin{equation}
\left. {x\tilde U_{RFI}^{Physical} \left( x \right)} \right|_{B.C.}  - \int {\tilde U_{RFI}^{Physical} \left( x \right)dx}  =  - \int {\tilde U_{RFI}^{Physical} \left( x \right)dx}.
\end{equation}

For vanishing $\tilde U_{RFI}^{Physical}(x)$ at the boundaries in (A.5), (A.3) is true for any $\tilde U_{RFI}^{Physical}(x)$.

\newpage

\section*{FIGURE CAPTIONS}

\textbf{Fig. 1:}  Inferred pdf profiles for the harmonic
oscillator obtained from the RFI framework (solid line) and the
FIM framework (dash-dots) evaluated from Eqs. (35) and (38),
respectively.  Here, $\omega^2$=0.5.
\\

\textbf{Fig. 2:}  Inferred pdf profiles for the quartic anharmonic
oscillator obtained from the RFI framework (solid line) and the
FIM framework (dash-dots) evaluated from Eqs. (35) and (38),
respectively.  Here, $\omega^2$=0.25 and $\varepsilon=0.125$.
\\
\newpage
\begin{figure}[thpb]
\centering
\begin{center}
\includegraphics[scale=1.00]{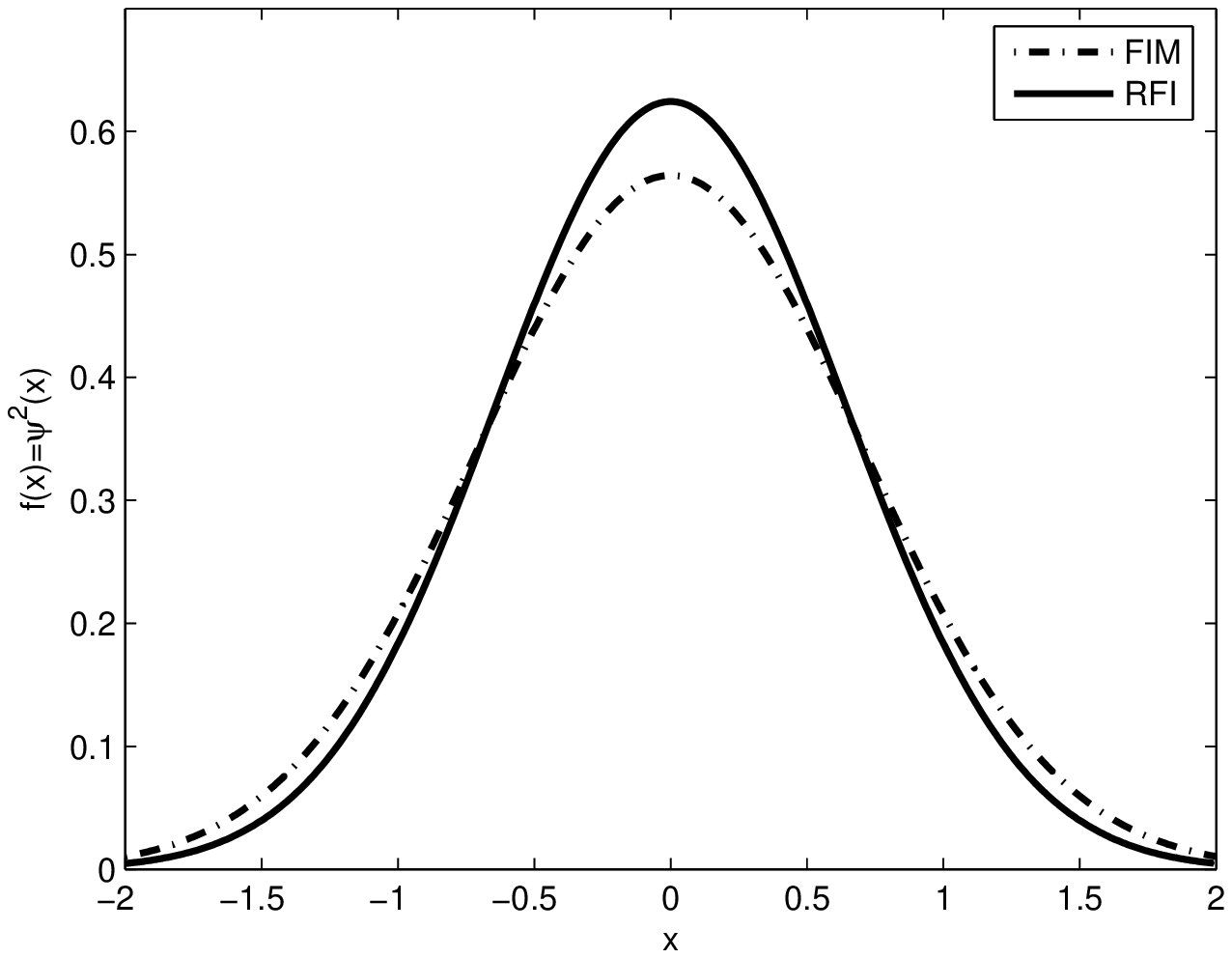}
\end{center}
\end{figure}

\begin{figure}[thpb]
\centering
\begin{center}
\includegraphics[scale=1.00]{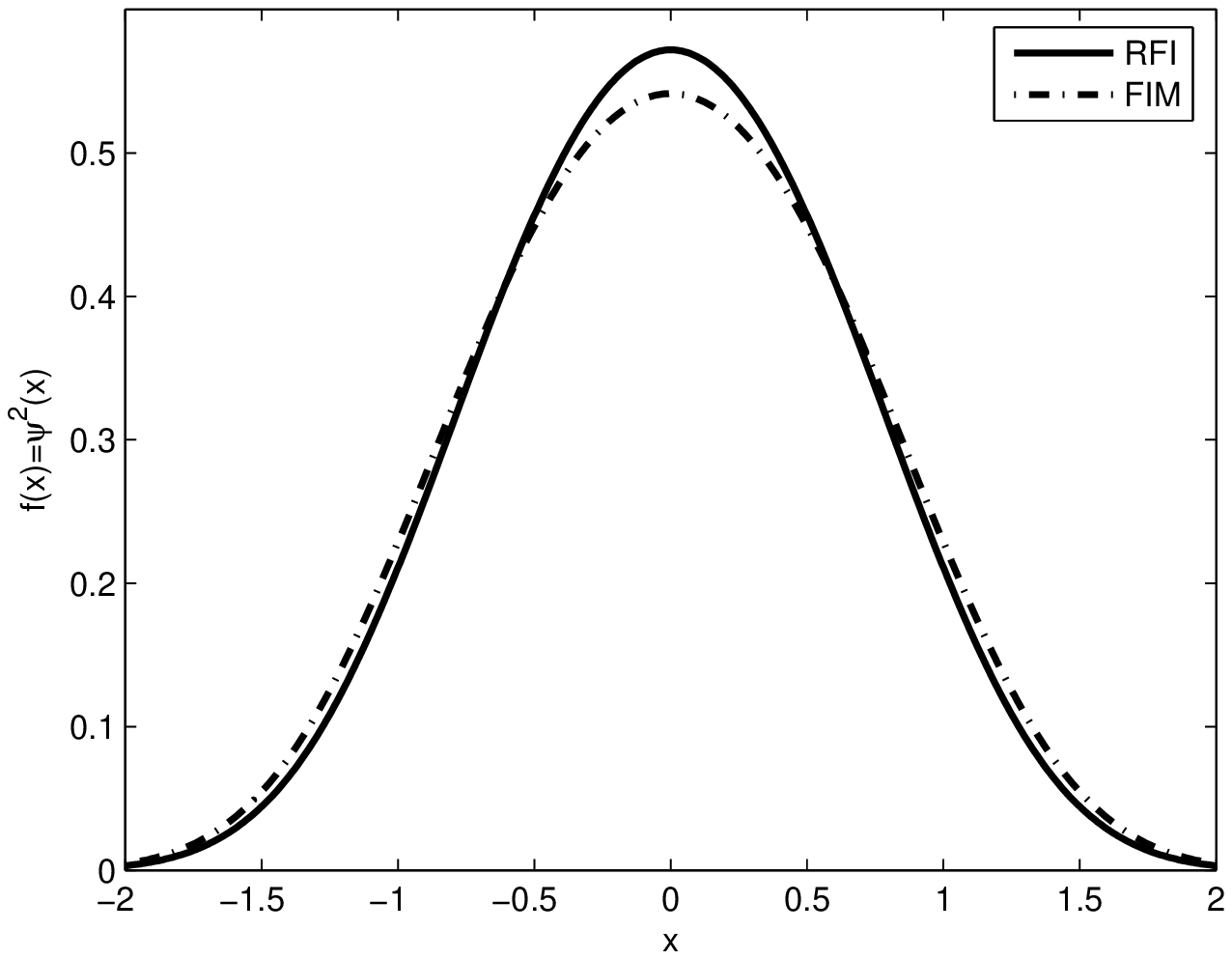}
\end{center}
\end{figure}

\end{document}